# Private and Secure Post-Quantum Verifiable Random Function with NIZK Proof and Ring-LWE Encryption in Blockchain

Bong Gon Kim[1], Dennis Wong[2], and Yoon Seok Yang[3]

[1]Department of Computer Science, Stony Brook University, New York, USA,

[2]Department of Computer Science, Macao Polytechnic University, Macao, China,

[3]Department of Computer Science, SUNY Korea University, Incheon, South Korea

ABSTRACT

We present a secure and private blockchain-based Verifiable Random Function (VRF) scheme addressing some limitations of classical VRF constructions. Given the imminent quantum computing adversarial scenario, conventional cryptographic methods face vulnerabilities. To enhance our VRF's secure randomness, we adopt post-quantum Ring-LWE encryption for synthesizing pseudo-random sequences. Considering computational costs and resultant on-chain gas costs, we suggest a bifurcated architecture for VRF design, optimizing interactions between on-chain and off-chain. Our approach employs a secure ring signature supported by NIZK proof and a delegated key generation method, inspired by the Chaum-Pedersen equality proof and the Fiat-Shamir Heuristic. Our VRF scheme integrates multi-party computation (MPC) with blockchain-based decentralized identifiers (DID), ensuring both security and randomness. We elucidate the security and privacy aspects of our VRF scheme, analyzing temporal and spatial complexities. We also approximate the entropy of the VRF scheme and detail its implementation in a Solidity contract. Also, we delineate a method for validating the VRF's proof, matching for the contexts requiring both randomness and verification. Conclusively, using the NIST SP800-22 of the statistical randomness test suite, our results exhibit a 98.86% pass rate over 11 test cases, with an average p-value of 0.5459 from 176 total tests.

## 1  Introduction

Many cryptographic protocols within blockchain architectures pivot fundamentally on inherent secure randomness. The Verifiable Random Function (VRF) [1] stands as an exemplar cryptographic element, ensuring the derivation of pseudo-random numbers distinguished by uniqueness, uniformly spread randomness, and accompanying verifiable proofs. Its extensive application spectrum, spanning lotteries [2], proof-of-stake systems [3–5], and protocols prioritizing privacy, foregrounds the imperative of crafting VRFs with true randomness.

The dynamics of the cryptographic domain accentuate the necessity for VRFs' adaptation as they become susceptible, especially to adversarial quantum computing attacks. The Shor's algorithm [6] for factorization looms as a significant quantum threat to conventional public key cryptographic paradigms. Anticipating the quantum adversarial attacks, it is incumbent to adopt post-quantum cryptographic advancements, with the spotlight on quantum defensibility and cryptographic resilience.

In this context, the Learning With Errors (LWE) problem [7] and its structured variant, Ring-LWE [8], have gained cryptographic prominence due to their innate quantum-attack resilience [9]. Lattice-based cryptographic frameworks, tracing back to Ajtai's hardness instances such as finding shortest vector problems [10], establish an influential cryptographic basis. Ring-LWE's introduction of structured noise and secrets within polynomial domains enhances efficiency while preserving robust security metrics [11].

Distinctively, Ring-LWE surpasses traditional lattice frameworks by streamlining computations and optimizing key dimensions, leveraging the algebraic intricacies of number-theoretic rings. In the emergent post-quantum horizon, where prevailing cryptographic infrastructures become susceptible against quantum-capable algorithms like Shor's [6], Ring-LWE's encryption mechanism inherently guarantees the safeguarding of encrypted content even amidst advanced quantum computation capabilities.

In decentralized frameworks like Ethereum and Algorand, MPC-based VRFs are crucial. For instance, Algorand extensively employs VRFs for *cryptographic self-selection*, maintaining *Selective Verifiers* (SV) and decision of block creation initiatives, highlighting the pivotal role of VRFs in preserving decentralized integrity.

Within the Ethereum area, VRFs are particularly central to consensus algorithms underpinning its decentralized fabric. Ethereum's transition from a Proof-of-Work to a Proof-of-Stake consensus protocol, as exemplified in the Ethereum 2.0 Beacon Chain [12], intensifies the significance of unpredictability in verifiable randomness. This form of randomness guarantees impartial, tamper-proof validator selections, imperative for block validations and attestations.

While Ring-LWE encryption's strength is favorable, it essentially introduces additional computational challenges. Particularly for blockchain ecosystems like Ethereum, where computational complexities have proportional *gas costs* [13], incorporating Ring-LWE could escalate transactional budget overheads. Arithmetic operations necessitated by Ring-LWE, attributable to its intricate polynomial nature, may culminate in elevated gas costs as compared to conventional cryptographic constructs [14].

Addressing these computational and cost challenges, off-chain computation emerges as a viable alternative. In the domain of MPC-driven VRFs, off-chain computations enable intricate cryptographic operations to offload outside the main net of blockchain, conserving on-chain resources and curbing associated gas costs.

Yet, off-chain computational approaches come with authenticity and validity concerns. Without authentic validation, malicious attackers such as Man-in-the-Middle (MIM) might inject spurious data, compromising system integrity [15]. Herein, the DID-based ring signature scheme [16] with accompanying NIZK proof, drawing inspiration from the Franklin-Zhang signature [17], bolstered by the Chaum-Pedersen proof [18] and the Fiat-Shamir heuristic [19], affirms the off-chain computation's integrity. These mechanisms vouch for the correctness of off-chain operations, ensuring adherence to expected outcomes and behaviors. Leveraging the DID-based ring signature scheme with NIZK proof, blockchain networks can ascertain the genuineness and precision of off-chain sourced data, fortifying decentralized systems' trust infrastructure.

Decentralized Identifiers (DID) [20] herald a Self-Sovereign-Identity (SSI) approach for identity authentication, underlining user-driven control and confidentiality. In MPC-based VRFs, where various nodes collectively generate random numbers, DIDs impart an augmented obfuscation layer. By obscuring the connection between real-world identities and their cryptographic counterparts, adversarial entities are thwarted from linking and discerning a specific node's contributory value. Thus, assimilating DIDs amplifies the VRF's defense mechanisms, augmenting system privacy [20]. Additionally, DIDs' interoperability across diverse platforms and ecosystems enhances their applicability, fostering a privacy-preserving milieu for VRF systems. This manuscript elucidates:

- A post-quantum VRF scheme, integrating DIDs and the ring signature scheme with NIZK proof, emphasizing off-chain validation.

- A smart contract-based VRF, harnessing MPC and Ring-LWE, influenced by Clercq et al.'s research [21].
- Detailed descriptions of our Solidity and Python implementations, algorithmic methodologies, and complexity evaluations (Fig. 2).
- Security argumentations, privacy explorations, and entropy assessments for Ring-LWE-based VRFs.
- NIST SP800-22 [22] outcomes, demonstrating a 98.86% success rate across 11 test cases and a mean $p$-value of 0.5459 from 176 evaluations (Figs. 4, 3, 5, and Table 1).

The organization of this paper is structured as follows: In Section 2, we provide a comprehensive exploration of VRF and Ring-LWE encryption to lay the foundational groundwork. Section 3 delineates our innovative VRF framework, elucidating its formal instantiation using NIZK proof, its distributed algorithmic features, and its seamless interfacing with smart contracts. We delve into the security dimensions in Section 4, accentuating the post-quantum adversarial advantage probability and crucial VRF key validations. In Section 6.1, we embark on a detailed analysis of the entropy characteristics of our MPC-based VRF, probing into the theoretical constraints of its randomness. Section 5 critically examines both the temporal and spatial complexities. The empirical results, derived from the NIST SP800-22 [22] test suite encompassing 11 test cases for evaluating randomness, are showcased in Section 6, alongside a depiction of our implementation using Solidity and Ganache. Section 7 offers a synthesis of our discussions and insights.

## 2 Preliminaries

### 2.1 Verifiable Random Function (VRF)

VRFs constitute a cornerstone for a plethora of cryptographic endeavors, emphasizing both unpredictability and verifiability. Resembling Pseudo-Random Functions (PRFs), VRFs distinguish themselves by an inherent ability: the generation of a proof for each of its outputs. This guarantees the verifier that a specific random number is obtained without disclosing the associated input [1]. Formally delineated, a VRF is structured around three polynomial-time algorithms:

- **KeyGen**: Produces a public key *PK* alongside a secret key *SK*.
- **Evaluate**: Given *SK* and input *x*, it yields an output *y* accompanied by a proof $\pi$.
- **Verify**: Utilizing *PK*, *x*, *y*, and $\pi$, it confirms the authenticity of *y*.

Micali et al.'s foundational VRF proposal [1] was predicated upon an RSA-based verifiable unpredictable signature method, achieving randomness through the Goldreich-Levin hardcore bit transformation [23, 24]. Subsequent breakthroughs, instantiated by the number-theoretical exponentiation-based PRF from Naor and Reingold [25], and seminal insights by Joux et al. on specific multiplicative groups [26] where a *Decisional Diffie-Hellman* problem becomes easy while the corresponding *Computational Diffie-Hellman* problem remains hard, have enriched the subsequent VRF research. The advent of bilinear pairing techniques [27] spurred further novel signature-oriented VRF designs such as Lysyanskaya's [28], which harnessed antecedent studies and the concept of an admissible hash function (AHF). The critical requisites for robust cryptographic capabilities in hash functions [29, 30] have invoked pursuits towards sculpting AHFs resilient under prevailing security paradigms [31].

Within the blockchain context, VRF's deterministically unpredictable property is instrumental in consensus mechanisms, fortifying fairness and obstructing potential adversary influence. Nonetheless, the emergent quantum computational age introduces vulnerabilities across classical cryptosystems. Considering VRFs frequently rely on challenges potentially vulnerable to quantum solutions, like the discrete logarithm problem [32], it's pivotal to anticipate and counteract these vulnerabilities.

Recent research in VRFs, due to Tibor et al., have introduced AHF architectures that bolster proof evaluation [33–36]. Our proposition takes a distinctive trajectory. By fusing a blockchain-oriented VRF through Multiparty Computation (MPC) with Ring-LWE encryption, our objective is to deliver augmented security, adaptability, and efficacy in random number synthesis for decentralized infrastructures.

## 2.2 Lattice-based Cryptography

A *lattice L* is systematically defined as a discrete subgroup within $\mathbb{R}^n$. Conceptually, it's the mesh resulting from assimilating all integer linear combinations of $n$ linearly independent vectors $\mathbf{b}_1, \mathbf{b}_2, \ldots, \mathbf{b}_n$ in $\mathbb{R}^n$. Such vectors compose the lattice's *basis*. Formally expressed as:

$$L(\mathbf{b}_1, \mathbf{b}_2, \ldots, \mathbf{b}_n) = \left\{ \sum_{i=1}^{n} x_i \mathbf{b}_i : x_i \in \mathbb{Z} \right\}$$

Cryptographically, the hardness of specific lattice challenges, particularly the Shortest Vector Problem (SVP) and the Learning With Errors (LWE) problem, underpins the security foundations of lattice cryptography [10, 37].

## 2.3 Quantum Robustness in Ring-LWE Encryption

Delving deeper into the lattice cryptographic sphere, the Ring-LWE emerges as a streamlined LWE variant, pivoting from vector realms to polynomial ring domains. This transition bestows Ring-LWE with computational efficiencies, rendering it fitting for cryptographic operations such as pseudo-random value generation and Verifiable Random Functions (VRF) [38].

Efforts synergizing Ring-LWE with random value generation have been profoundly influential. As an exemplar, Abraham's research elucidates a potent VRF mechanism propelled by Ring-LWE, emphasizing both agility and fortification [39].

Peikert's research in 2010 shed light on the quantum-resistant properties of LWE, and consequently, Ring-LWE [40]. This also resonates in subsequent research, notably from Regev and Lyubashevsky, affirming Ring-LWE's impermeability against quantum-hostile maneuvers [41, 42].

## 2.4 Chaum-Pedersen Zero Knowledge Logarithm Equality Proof

The Chaum-Pedersen logarithm equality proof [18] allows a prover to assert "I am aware of $x$ such that $h_1 = g_1^x$ and $h_2 = g_2^x$", without disclosing $x$. Given:

- $g_1, g_2$: Public group generators.
- $h_1, h_2$: Asserted values where $h_1 = g_1^x$ and $h_2 = g_2^x$.

  The proof operates in the following manner:

1. Prover picks a random $r$ and communicates commitments $t_1 = g_1^r$ and $t_2 = g_2^r$ to the verifier.
2. Verifier replies with a random challenge $c$.
3. Prover calculates $s = r + c \times x$ and conveys it.
4. Verifier checks the relations $t_1 = g_1^s \times h_1^{-c}$ and $t_2 = g_2^s \times h_2^{-c}$.

## 2.5 Fiat-Shamir Transformation

The Fiat-Shamir heuristic [19] converts an interactive zero-knowledge protocol to a non-interactive version. From an initial commitment and subsequent interaction, the prover derives the challenge via a deterministic function, generally a cryptographic hash.

To render the Chaum-Pedersen proof non-interactive:

1. Instead of a verifier's challenge $c$, calculate $c$ as $c = \text{Hash}(t_1 \parallel t_2 \parallel \text{relevant\_public\_information})$.
2. Choose a random $r$.
3. Determine commitments $t_1, t_2$.
4. Evaluate $c = \text{Hash}(t_1 || t_2 || \text{additional\_data})$.
5. Determine $s$.
6. Proof $\pi$ is represented as $(t_1, t_2, s)$.

Proof verification proceeds as follows: Given $\pi = (t_1, t_2, s)$ and associated public information:

1. Determine $c = \text{Hash}(t_1 || t_2 || \text{additional\_data})$.
2. Confirm relations for $t_1$ and $t_2$ as described in Section 2.4.

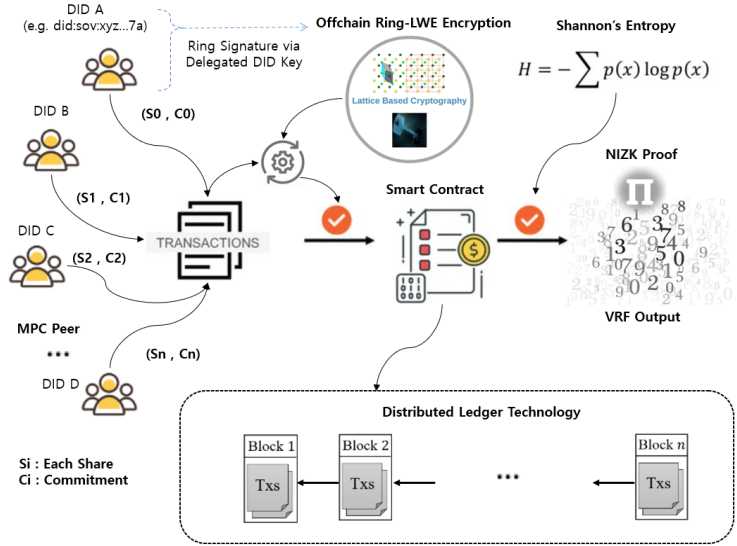

**Fig. 1.** A quantum-secure VRF system incorporating MPC and DIDs, where off-chain Ring-LWE encryption is synthesized with the DID-based ring signature scheme via DKG and NIZK proof.

---

**Algorithm 1** Core VRF computation process

**Require:** $roundId \in \mathbb{Z}_{\geq 0}$
**Ensure:** None
1: **Require:** $combinedBlockHash$ of 3 recent blocks $> 0$
2: DID Keys $sk_i, pk_i, sk'_i, pk'_i \leftarrow \text{GenerateKeys}$
3: Encrypted key $sk_{enc} \leftarrow \text{DelegateKey}(sk'_i)$
4: $weightedSum \leftarrow \text{MultiplyAndAdd}(commitment, share)$
5: $seed \leftarrow \text{EncodeData}(block, msg.sender, weightedSum)$
6: **Emit:** $\text{OnchainSeedReady}(seed, msg.sender, R, sk_{enc})$
7: **Off-chain:**
8: Cipher texts $(c1, c2) \leftarrow \text{rlwe\_processing}(seed)$
9: $VRF_{\text{output}} \leftarrow \text{keccak256}(c1, c2)$
10: $sk'_i \leftarrow \text{Decrypt}(sk_{off}, sk_{enc})$
11: signature $\sigma \leftarrow \text{RingSign}(sk', seed, VRF_{output}, R)$
12: NIZK proof $\pi \leftarrow \text{OffchainSign}(sk', seed, VRF_{output}, R)$
13: **Off-chain to On-chain:** $\text{Submit\_To\_Blockchain}(\pi)$
14: **On-chain:** $\text{VerifySign}(\pi, R)$
15: **Emit:** $\text{ComputationFinished}(roundId, \pi)$

# 3 Proposed VRF System

## 3.1 System Architecture

We propose an advanced VRF scheme tailored for blockchain platforms, incorporating DIDs and post-quantum secure Ring-LWE encryption [21]. This integrates the DID-based ring signature method [16] and NIZK proof, utilizing a delegated key generation approach. Our system leverages lattice cryptography for random number derivation via off-chain from seed inputs, collaboratively produced on-chain. The comprehensive VRF system is illustrated in Fig. 1. A rigorous analysis of security and entropy properties is presented in Sections 4 and 6.1.

In Fig. 1, we delineate a high-level architecture of our DID-based VRF mechanism employing Ring-LWE encryption, highlighting primary protocols and related solutions. The procedure begins with decentralized MPC-based seed generation on a smart contract. Each participant contributes a share and a corresponding commitment to generate this seed, efficiently computed within Solidity. Once the seed is set, an off-chain blockchain monitor, leveraging the Web3 protocol, is notified and triggers the `RLWE_enc2()` function to apply Ring-LWE encryption. The resulting ciphertexts $c1$ and $c2$ are relayed to the smart contract through the `submitRLWEResult` function.

To ensure the validity of off-chain Ring-LWE encryption, we adopted a ring signature scheme, inspired by Franklin-Zhang [17], enhanced by the Fiat-Shamir heuristic [19] for a non-interactive commit protocol and the Chaum-Pedersen's method [18] for NIZK proof generation. A challenge arose due to the fact that the ring group doesn't exist in the off-chain, while it exists in the on-chain facilitating the MPC-based seed generation.

We solved this problem via a delegated key generation protocol. During the on-chain group setup for MPC and ring signature, each participant $i$ generates dual key pairs: $(sk_i, pk_i)$ and $(sk'_i, pk'_i)$. The former serves on-chain operations, the latter acts as a delegation instrument. Both keys are committed to the contract. For MPC seed derivation, $sk_i$ is deployed. After seed generation, a designated participant encrypts their $sk'_i$ using the off-chain party's public key, $pk_{\text{off}}$, and dispatches the encrypted result to the contract. Upon this trigger, the off-chain module deciphers the content, recovering $sk'_i$.

Upon the acquisition of the delegated secret key, the off-chain module effectively impersonates participant $i$, crafting the ring signature. Using $sk'_i$, the off-chain module signs, amalgamating public keys from on-chain contributors, inclusive of $pk'_i$. The NIZK proof, attesting to signature validity, is constructed using the Fiat-Shamir heuristic [19] alongside Chaum-Pedersen's proof [18]. External parties can then validate the ring signature with the public key set, verifying the off-chain computation's authenticity.

The algorithm, detailed in Algorithm 1, delineates the core computational flow of our VRF system, centered on a confidential DID-based MPC seed creation and Ring-LWE encryption.

## 3.2 VRF Formal Instantiation:

Given our MPC-based Ring-LWE VRF system, we formalize the functions as follows:

1. **Key Generation:** Let $\mathcal{D}$ be the domain of all possible security parameters. Then, the function Gen is represented as:

   $$\text{Gen} : \mathcal{D} \to \mathcal{K} \times \mathcal{K}, \qquad \textit{where } \text{Gen}(1^\lambda) = (PK, SK)$$

   for $\lambda \in \mathcal{D}$, and $(PK, SK)$ are the public and private key pairs for participants, respectively. More details are covered in Section 3.4.

2. **Evaluate:** Let $\mathcal{S}, \mathcal{V}$ be the domains of all possible seeds and VRF system output of the tuple $(VRF_{output}, Proof_\pi)$. Then, the function Eval is represented as:

$$\text{Eval} : \mathcal{S} \to \mathcal{V}, \qquad where \text{ Eval}(Seed) = (VRF_{output}, \pi)$$

for $Seed \in \mathcal{S}$, and $VRF_{output}$ is the generated output from the Ring-LWE encryption. More details are covered in Section 3.3, 3.4.

3. **Verify:** Let $\mathcal{P}$ be the domain of all possible VRF proofs. Then, the function Ver is represented as:

$$\text{Ver} : \mathcal{P} \to \{\texttt{TRUE}, \texttt{FALSE}\}, \qquad where \text{ Ver}(\pi) = \texttt{TRUE}$$

if and only if the NIZK proof $\pi$ is valid. More details are covered in Section 3.3, 3.4.

---

**Algorithm 2** DID-based Ring Signature Scheme
---
1: **procedure** RINGSIGN($sk'_i, Seed, VRF_{output}, R$)  ▷ $R$ is the set of DID public keys
2:     Select random value $t$
3:     Compute $T = g^t$
4:     **for** $i = 2$ to $n$ **do**
5:         Select random values $c_i, s_i$
6:         Compute $e_i = g^{s_i} \times Y_i^{c_i}$
7:     **end for**
8:     Compute $c' = H(VRF_{\text{output}}, Seed, T, e_2, \ldots, e_n)$
9:     Compute $c_1$ such that: $c_1 + c_2 + \cdots + c_n = c'$
10:    Compute $s_1 = t - sk'_i \times c_1$
11:    $\sigma = (T, c_1, c_2, \ldots, c_n, s_1, s_2, \ldots, s_n)$
12:    **return** $\sigma$
13: **end procedure**
14: **procedure** VERIFYSIGN($\pi, R$)
15:    Parse $\pi$ as $(VRF_{\text{output}}, Seed, sigma)$
16:    Parse $\sigma$ as $(T, c_1, c_2, \ldots, c_n, s_1, s_2, \ldots, s_n)$
17:    **for** $i = 1$ to $n$ **do**
18:        Compute $e_i = g^{s_i} \times Y_i^{c_i}$
19:    **end for**
20:    Compute $c' = H(VRF_{\text{output}}, Seed, T, e_2, \ldots, e_n)$
21:    **if** $c' = c_1 + c_2 + \cdots + c_n$ **then**
22:        **return** True
23:    **else**
24:        **return** False
25:    **end if**
26: **end procedure**

---

### 3.3 Ring Signatures Integrated with VRF

In Algorithm 2, we introduce a DID-based ring signature scheme [16] that provides ambiguous signer identification, allowing verification of a member's signature without pinpointing the exact signer. This ensures inherent anonymity against passive adversaries and ensures adaptive security under the random oracle model.

**Setup** Given a ring $R$ consisting of participants $\mathcal{P} = \{P_1, P_2, \ldots, P_n\}$ where each $P_i$ possesses a private key $x_i$ and its public counterpart $Y_i = g^{x_i}$. Assuming $c1$ and $c2$ as RLWE encryption outputs and $g$ as a prime cyclic group $G$'s generator. The VRF evaluation function is defined as:

$$F_{eval} : \text{MPC-Based-Seed} \to \text{keccak256}(c1, c2) \tag{1}$$

**Signature Creation by $P_1$** When $P_1$ aims to incorporate the $VRF_{output}$ into the signature:

1. Compute $T$ using a random $t$: $T = g^t$.
2. For each $i$ from 2 to $n$, calculate $e_i$ with randomly chosen $c_i$ and $s_i$.
3. Derive hash $c'$ via: $c' = H(VRF_{output}, Seed, T, e_2, \ldots, e_n)$.
4. Determine $c_1$ and $s_1$ with respect to $c'$.
5. Construct ring signature $\sigma$ and its proof $\pi$.

The proof $\pi$ signifies that the VRF output, $VRF_{output}$, endorsed by an exclusive ring $R$ member, is tied to seed value $Seed$ and attested by signature $\sigma$.

**Proof Validation** The VRF verification function is:

$$F_{ver} : \pi \to \text{TRUE or FALSE} \qquad (2)$$

To validate the proof $\pi$ related to signature $\sigma$ and $VRF_{output}$:

1. For each $i$ in 1 to $n$, derive $e_i$.
2. Calculate hash $c'$.
3. Ascertain the signature's validity based on $P(c')$.

**Correctness Proof** Let's denote:

- $\Sigma$ as signature generation.
- $\Phi$ as signature verification.

The objective is to assert that given any Seed and VRF output, a validly constructed signature ensures a successful verification:

$$\forall \text{ Seed}, VRF_{output}, \Sigma(\text{Seed}, VRF_{output}) \implies \Phi(\pi)$$

1. Assume: $\Sigma(\text{Seed}, VRF_{output})$, i.e., the signature generation protocol $P_1$ is duly executed.
2. From $\Sigma$, it follows:
   $c' = H(VRF_{output}, Seed, T, e_2, \ldots, e_n)$ and
   $c_1 + c_2 + \cdots + c_n = c'$
3. In $\Phi$, $c'$ is recalculated to ascertain $c' = c_1 + c_2 + \cdots + c_n$.
4. With $H$ being collision-resistant, distinct inputs yield unique outputs. Therefore, a valid $\Phi$ indicates a genuine $\Sigma$.
5. Conclusion: $\Sigma(\text{Seed}, VRF_{output}) \implies \Phi(\pi)$.
   This confirms the signature's correctness.

### 3.4 Delegated Key Generation (DKG)

We describe a method enabling an off-chain entity to generate a DID-based ring signature [16] on behalf of an on-chain user without exposing the user's DID secret key.

**Protocol**

i. Each participant $i$ constructs two DID key pairs: $(sk_i, pk_i)$ for on-chain tasks and $(sk'_i, pk'_i)$ as a delegation key, as outlined in Algorithm 3.
ii. Participants commit both $pk_i$ and $pk'_i$ to the contract.
iii. For the MPC seed creation, participants employ $sk_i$.
iv. After MPC seed extraction, a selected participant encrypts $sk'_i$ using the off-chain component's public key, $pk_{off}$, and commits it to the contract.
v. The off-chain unit, upon notification, decrypts the value with its private key, $sk_{off}$, retrieving $sk'_i$ to produce the ring signature.

**Algorithm 3** Delegated Key Generation

```
 1: procedure GENERATEKEYS(∅)
 2:     (sk_i, pk_i) ← KEYGEN
 3:     (sk'_i, pk'_i) ← KEYGEN
 4:     return (sk_i, pk_i, sk'_i, pk'_i)
 5: end procedure
 6: procedure DELEGATEKEY(sk'_i)
 7:     encryptedKey ← ENCRYPT(sk_off, sk'_i)
 8:     return encryptedKey
 9: end procedure
10: procedure OFFCHAINSIGN(sk_enc, Seed, VRF_output, R)     ▷ R is the set of DID public keys
11:     sk'_i ← DECRYPT(sk_off, sk_enc)
12:     sigma ← RINGSIGN(sk'_i, Seed, VRF_output, R)
13:     π ← (VRF_output, Seed, sigma)
14:     return π
15: end procedure
```

**Ring Signature via Delegation Protocol** With the delegated $sk'_i$, the off-chain unit initiates the DID ring signature as if it were participant $i$, as detailed in Algorithm 3.

i. With $sk'_i$, the off-chain unit signs and aggregates public keys of on-chain users, including $pk'_i$.
ii. The signature's authenticity is ensured through a non-interactive zero-knowledge (NIZK) proof via the Fiat-Shamir heuristic melded with Chaum-Pedersen proof.

**Verification** Observers, using Algorithm 2, can autonomously validate the ring signature against the aggregated DID public keys, including $pk'_i$, affirming the off-chain process's validity and integrity.

**Algorithm 4** Off-chain Blockchain Listener for Ring-LWE Computation

```
 1: function HANDLE_EVENT(event)
 2:     sumHash ← event['args']['sumHash']
 3:     c1, c2 ← rlwe_processing(sumHash)
 4:     tx_hash ← contract.functions.submitRLWEResult(c1, c2)
                .transact()
 5:     tx_receipt ← w3.eth.waitForTransactionReceipt(tx_hash)
 6:     if tx_receipt.status == 1 then
 7:         print("Result and proof submitted successfully!")
 8:     else
 9:         print("Submission failed.")
10:     end if
11: end function
12: function BLOCKCHAIN_LISTENER
13:     w3 ← Web3(Web3.HTTPProvider
                ('http://localhost:8545'))
14:     comment: Assuming contract ABI and address are available
15:     contract ← w3.eth.contract
           (address = contract_address, abi = contract_abi)
16:     event_filter ← contract.events.OnchainMpcSeedReady
                .createFilter(fromBlock = 'latest')
17:     while True do
18:         for all event in event_filter.get_new_entries() do handle_event(event)
19:         end for
20:     end while
21: end function
```

### 3.5 Off-chain Ring-LWE Computation

The off-chain Ring-LWE computation is encapsulated within the blockchain event listener designed for Ethereum, as delineated in Algorithm 4. For deployment, we opt for Ganache[1]. This listener anticipates the `OnchainMpcSeedReady` event, subsequently triggering the Ring-LWE encryption mechanism. Key operations are:

1. **RLWE Execution (`rlwe_processing` function):** Engages an external binary (`LWE.exe`) for Ring-LWE encryption. Upon receiving a seed (`sumHash`), the binary is activated and its output seized.
    – Error-handling ensures smooth encryption.
    – The encrypted output is extracted post-binary execution.
2. **Event Management (`handle_event` function):** Post event-detection, this manages ensuing actions, primarily:
    – Deriving the `sumHash` from the event.
    – Utilizing `sumHash` for RLWE encryption.
    – Relaying the encryption, denoted as `c1` and `c2`, back to the Ethereum contract and verifying its successful transmission.
3. **Ethereum Network Connection (`blockchain_listener` function):** Establishes the Ethereum connection via Web3 in Python. Detailed steps include:
    – Establishing a Web3 connection to an Ethereum node.
    – Configuring the smart contract using its ABI and address.
    – Continuously tracking the `OnchainMpcSeedReady` event and, when detected, invoking `handle_event`.

The blockchain listener is adeptly triggered via the on-chain event and ensures off-chain Ring-LWE cryptographic computations and the corresponding return of the encrypted outputs.

### 3.6 Ring-LWE Encryption: `RLWE_enc2()`

The `RLWE_enc2()` function, presented in Algorithm 5 and inspired by [21], realizes the encryption mechanism of the Ring-LWE. Given a message, public key, and value polynomials denoted as $m(x)$, $a(x)$, and $p(x)$ respectively, all within the polynomial ring $R$, and leveraging the Number Theoretic Transform (NTT) from [43], the encryption procedure is summarized as:

1. Encode $m(x)$ by scaling with $\frac{Q}{2}$, denoted as $encoded\_m(x)$ where $Q$ is a system parameter.
2. Sample three error polynomials $e_1(x)$, $e_2(x)$, and $e_3(x)$ from a predefined error distribution.
3. Update $e_3(x)$ as $e_3(x) = e_3(x) + encoded\_m(x)$.
4. Transform $e_1(x)$, $e_2(x)$, and $e_3(x)$ into the NTT domain.
5. Determine the ciphertext polynomials as:

$$c_1(x) = e_2(x) + a(x) \cdot e_1(x)$$
$$c_2(x) = e_3(x) + p(x) \cdot e_1(x)$$

6. Organize coefficients of $c_1(x)$ and $c_2(x)$ for transmission.

---

[1] https://trufflesuite.com/ganache/

**Algorithm 5** RLWE_enc2 Encryption Algorithm
---
1: **procedure** RLWE_PROCESSING(*seed*)
2:    **comment:** Assuming LWE.exe takes seed value as a command-line argument
3:    *result* ← subprocess.run(["./*LWE.exe*", str(*seed*)])
4:    **if** *result*.returncode ≠ 0 **then**
5:       **raise Exception**(f"RLWE Encryption failed with error: result.stderr")
6:    **end if**
7:    $c1, c2 \leftarrow result.stdout.\text{strip}()()$
8:    **return** Ciphertexts $c1, c2$
9: **end procedure**
10: **procedure** RLWE_ENC2(a, c1, c2, m, p)
11:    encoded_m ← $m \times \frac{Q}{2}$
12:    $e_1, e_2, e_3 \leftarrow$ knuth_yao2() × 3
13:    $e_3 \leftarrow e_3 +$ encoded_m
14:    $e_1, e_2, e_3 \leftarrow$ fwd_ntt2($e_1$), fwd_ntt2($e_2$), fwd_ntt2($e_3$)
15:    $c_1 \leftarrow e_2 + a \times e_1$
16:    $c_2 \leftarrow e_3 + p \times e_1$
17:    $c_1, c_2 \leftarrow$ rearrange2($c_1$), rearrange2($c_2$)
18: **end procedure**

## 4 Security and Privacy

### 4.1 Assumptions

Let $\lambda$ be the security parameter, while $\epsilon(\lambda)$ denotes a negligible function relative to $\lambda$. The function *Gen*($1^\lambda$) generates distinct DID keys for participants. $Pr[\cdot]$ denotes an event's likelihood and $\mathbb{M}$ is the space of input messages.

- RING-LWE POSTULATION (QUANTUM RESILIENT): Considering quantum machines can efficiently tackle discrete logarithm and integer factorization, but not the Ring-LWE, we infer the Ring Learning With Errors challenge remains quantum-resistant.
- QUANTUM RANDOM ORACLE HYPOTHESIS (QROM): Hash functions, for instance, `keccak256()`, are deemed quantum-proof and function as a quantum random oracle, meaning unforeseen outcomes remain unpredictable even to quantum requests.
- INTEGRITY OF PRIVATE RING SIGNATURES: Any polynomial-time opponent is incapable of producing a legitimate distinct ring signature devoid of the private key.
- MPC INTEGRITY: Predicting or altering the concluding MPC result is computationally daunting for adversaries unless they dominate a majority of the members.

### 4.2 Formal VRF Requirements

**Uniqueness** Within a VRF framework, *uniqueness* asserts that a consistent output is generated for any particular input.

    **Proof**: Designate $\Phi$ as our VRF mechanism. Given an input *m* (our MPC seed), and the DID key set ($DID_{sk}, DID_{pk}$):

$$\forall m_i, m_j \in \mathbb{M} \ (i \neq j), \exists! \ \text{VRF}_{\text{output}} \text{ such that } \Phi(m_i, DID_{sk}) = \Phi(m_j, DID_{sk}) = \text{VRF}_{\text{output}}$$

**Verifiability** For VRF, verifiability guarantees a verifier can authenticate an output *y* and its associated proof $\pi$ for an input *x*.

    **Proof**: Given Input *m* (MPC seed), VRF outcome *y*, Proof $\pi$, and DID keys $DID_{sk}$ and $DID_{pk}$.

If the prover generates $y$ and $\pi$ using $m$ and $DID_{sk}$:

$$y, \pi = \Phi(m, DID_{sk})$$

The verifier, applying $DID_{pk}$, can ascertain $y$ as the legitimate VRF result for $m$:

$$\forall m \in \mathbb{M}, y, DID_{pk}, \pi, \ \Phi(m, DID_{pk}, \pi) \Rightarrow y = \text{VRF}_{\text{output}}(m)$$

This indicates a consistent outcome when the verifier uses the public key and proof to authenticate the input.

**Randomness** For a VRF system, randomness ensures that the output appears random and unpredictable. Given a VRF output $y$ for an input $x$, without the associated proof $\pi$, one cannot distinguish $y$ from a random value.

**Proof**: Given an Input $m$, VRF output $y$, Proof $\pi$, Adversary $\mathcal{A}$ trying to distinguish $y$ from a random value, without the proof $\pi$, the advantage Adv of $\mathcal{A}$ in distinguishing $y$ from a random value is negligible:

$$\forall m \in \mathbb{M}, y, \ \text{without } \pi, \ \text{Adv}(\mathcal{A}(y)) \leq \epsilon(\lambda)$$

We've demonstrated that the MPC-based Ring-LWE VRF system integrated with the DID-based ring signature and NIZK proof satisfies the three pivotal properties of a VRF: uniqueness, verifiability, and randomness. These properties, complemented by the security guarantees from our previous discussions, affirm that the VRF system is robust and secure.

### 4.3 MPC-based Seed Integrity

The integrity of the MPC-based seed relies on the commitments made by participants. Under the assumption of a random oracle model for the hash function, the probability that an adversary can produce a commitment for a value without knowing that value is negligible. Formally:

$$\Pr[Seed' \leftarrow \mathcal{A}(\text{Commitments}) : Seed' = Seed] \leq \epsilon(\lambda)$$

### 4.4 Unforgeability under Chosen Message Attack (CMA)

For all messages $m_1, m_2, \ldots, m_k$ chosen adaptively by $\mathcal{A}$, where signatures $\sigma_1, \sigma_2, \ldots, \sigma_k$ of VRF outputs are produced, the probability that $\mathcal{A}$ produces a new valid signature $\sigma^*$ for a new message $m^*$ without knowledge of the DID private key that signed the DID-based ring signature is negligible.

$$\Pr[\sigma^* \leftarrow \mathcal{A}(m_1, \sigma_1, \ldots, m_k, \sigma_k) :$$
$$\Phi(m^*, \sigma^*) = \text{True} \wedge m^* \notin \{m_1, \ldots, m_k\}] \leq \epsilon(\lambda)$$

### 4.5 Post-Quantum Security

**Definitions**

- Let $RLWE_{q,\chi}$ be the Ring-LWE problem with modulus $q$ and error distribution $\chi$.
- $\mathcal{A}_{RLWE}$ is a polynomial-time adversary $\mathcal{A}$ trying to solve the Ring-LWE problem.
- $\mathcal{A}_{SVP}$ is an adversary trying to solve the approximate SVP in ideal lattices.
- $\alpha$ is the approximation factor for the SVP problem.

**Ring-LWE Problem** Given a random polynomial $a$ from a ring $R_q$ and a "noisy" product $b = (a \times s) + e \mod q$ where $s$ is a secret polynomial and $e$ is an error polynomial drawn from $\chi$, the goal is to recover $s$ or distinguish $b$ from a random polynomial.

**Security Proof for RLWE_enc2**

i. RING-LWE ASSUMPTION: It's computationally infeasible for a polynomial-time quantum or classical adversary to solve the Ring-LWE problem or distinguish between a valid Ring-LWE sample and a random one.
∀ Ring-LWE samples $s$,

$$Pr[s' \leftarrow \mathcal{A}_{RLWE}(a,b) : s' = s] \leq \epsilon(\lambda)$$

ii. REDUCTION TO SVP: If there exists a polynomial-time algorithm $\mathcal{A}_{RLWE}$ that can solve the $RLWE_{q,\chi}$ problem, then there exists an algorithm $\mathcal{A}_{SVP}$ that can solve the $\alpha$-approximate SVP in ideal lattices in polynomial time. ∀ lattices derived from Ring-LWE samples,

$$Pr[v' \leftarrow \mathcal{A}_{SVP}(Lattice) : ||v'|| \leq \alpha \times ||v_{shortest}||]$$

$$\geq Pr[s' \leftarrow \mathcal{A}_{RLWE}(a,b) : s' = s]$$

Here, $v_{shortest}$ is the shortest non-zero vector in the lattice. The $\alpha$-approximate SVP requires finding a vector whose length is within $\alpha$ times the shortest vector.

iii. POST-QUANTUM SECURITY: Given that SVP in ideal lattices is believed to be hard for quantum computers (there's no known polynomial-time quantum algorithm for this problem), the security of Ring-LWE and, in turn, `RLWE_enc2()` remains even in the presence of quantum adversaries.
∀ quantum adversary queries to $RLWE_{q,\chi}$,

$$Pr[s' \leftarrow \mathcal{A}_{RLWE}(QuantumQueries) : s' = s] \leq \epsilon(\lambda)$$

Overall, the security of the `RLWE_enc2()` function, as used in our VRF system, hinges upon the hardness of the Ring-LWE problem, which can be reduced to the hardness of the SVP in ideal lattices. This provides assurance of the post-quantum security of the function.

### 4.6 DKG Security

- **Confidentiality**: Employing asymmetric encryption ensures that $sk'_i$ remains confidential on-chain. Only the off-chain component, possessing $sk_{\text{off}}$, can decrypt this.
- **Integrity**: The ring signature confirms the integrity of the computation executed by the off-chain component.
- **Redundancy**: For backup, multiple participants might delegate their keys. This allows the off-chain component to choose from any of the provided keys, should one be unavailable.
- **Revocation**: A key pair update mechanism allows any participant to revoke or replace their delegation key pair, if they suspect potential misuse.
- **Non-repudiation**: The utilization of a specific delegated key for the ring signature holds the corresponding participant accountable, ensuring they cannot repudiate their involvement.

### 4.7 DID Privacy (GDPR Compliance)

By using DIDs, we ensure that every participant has a self-sovereign identity which enhances privacy. With unique DIDs, the system ensures that participants can prove their identity without revealing any personal data. Our DID-based private ring signature scheme in Section 3.3 further ensures that even when a participant signs, their specific identity remains hidden among the members of the ring. DIDs, as a decentralized identity, provide users control over their identity without relying on centralized authorities. In this context, DIDs are used for verification rather than identification, ensuring participants' actions are verifiable without revealing their exact identities.

Given the DID-based ring signature and QROM, the ability of any adversary (including quantum adversaries) to link a signature to a specific DID or to single out any individual signer becomes negligible. Thus, the system satisfies GDPR[2] requirements in terms of unlinkability, inference protection, and prevention of singling-out. Formally: ∀ DID in the ring,

$$Pr[\text{DID}^* \leftarrow \mathcal{A}(\sigma) : \text{DID}^* \text{ is the actual signer}] \leq \epsilon(\lambda)$$

## 5  Complexity Analysis

In this section, we provide an in-depth complexity analysis of the proposed MPC-based Ring-LWE VRF system. We break down the major components and evaluate both their time and space complexities, providing insights into the system's efficiency.

### 5.1  Temporal Complexity

- **MPC Seed Generation:** Predominantly driven by hashing, with complexity $\Theta(k)$ per operation. Basic arithmetic tasks run in $\Theta(1)$, culminating in a total complexity of $\Theta(n + k)$ for $n$ participants.
- **RLWE_enc2() Function:** The primary factor is polynomial multiplication with complexity $\Theta(M \log M)$. Accompanying operations like NTT and rearrangements align with this complexity.
- **submitRLWEResult() Function:** Centralized around hashing, the complexity is $\Theta(k)$.
- **Signature Generation:** Exponentiation tasks, notably $T$, exhibit a complexity of $\Theta(\log p)$, where $p$ is the group modulus. Including iterations across participants, the complexity scales to $\Theta(n \times \log p)$.
- **Total Time Complexity:** Summing up the components yields:

$$\Theta(n + 2k + M \log M + (n + 1) \log p)$$

### 5.2  Space Complexity

The space complexity of cryptographic systems is largely influenced by the data structures and variables used during computation.

- **MPC Seed Generation:** Given its hashing nature, it would require space proportional to the number of participants, i.e., $\Theta(n)$.
- **RLWE_enc2() Function:** Being polynomial-based, it demands space proportional to the polynomial degree, leading to a space complexity of $\Theta(M)$.
- **submitRLWEResult() Function:** This function, being related to hashing and data verification, would require a space complexity of $\Theta(k)$, which is the output size of the hash.
- **DID-based Ring Signature Generation:** The dominant space overhead here comes from the storage of random values $c_i$ and $s_i$ for all participants. Thus, its space complexity is $\Theta(n)$. The storage of the exponentiated values, such as $T$, is constant, adding only a fixed overhead.
- **Aggregate Space Complexity:** The combined space complexity, taking into account all the components mentioned, can be articulated as:

$$\Theta(n + k + M)$$

---

[2] https://gdpr-info.eu/recitals/no-26/

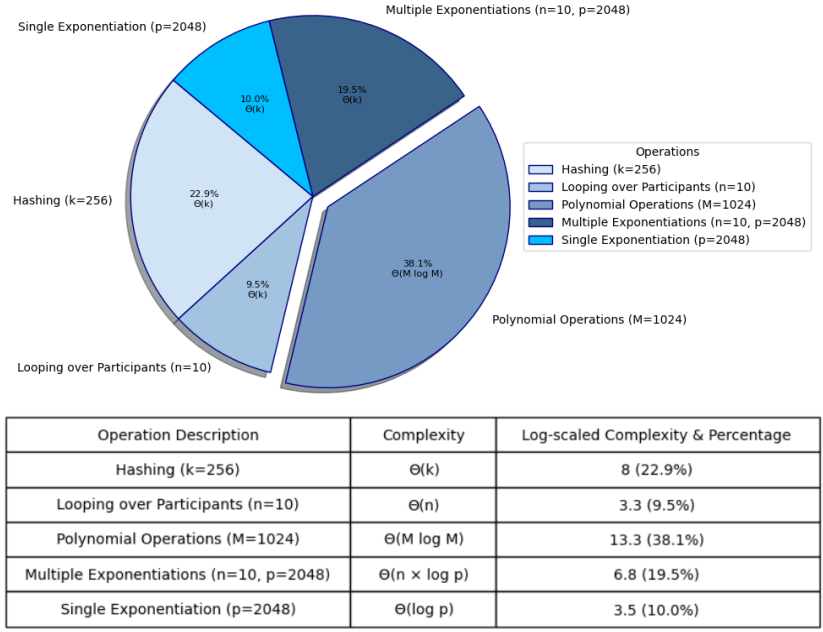

| Operation Description | Complexity | Log-scaled Complexity & Percentage |
|---|---|---|
| Hashing (k=256) | Θ(k) | 8 (22.9%) |
| Looping over Participants (n=10) | Θ(n) | 3.3 (9.5%) |
| Polynomial Operations (M=1024) | Θ(M log M) | 13.3 (38.1%) |
| Multiple Exponentiations (n=10, p=2048) | Θ(n × log p) | 6.8 (19.5%) |
| Single Exponentiation (p=2048) | Θ(log p) | 3.5 (10.0%) |

**Fig. 2.** Log-scaled Complexity Contributions in VRF System.

### 5.3 Efficiency Approximation

Our assessment underscores that the efficiency of the MPC-based Ring-LWE VRF system largely depends on both the number of participants, $n$, and the polynomial degree, $M$. Although constants $k$ and $p$ do affect the system, their impact remains static. For peak performance, adjusting the values of $n$ and $M$ based on the computational power of the hosting environment is pivotal, especially for high-frequency use cases or resource-limited settings.

For the sake of delineating dominant complexities as outlined in Sections 5.1 and 5.2, we assign typical values to our parameters: $k$, $n$, $M$, and $p$. Note that these are approximations and can differ in practical applications.

- $k$ (hash size) = 256 (common in cryptographic hashes)
- $n$ (number of participants) = 10
- $M$ (polynomial degree) = 1024 (standard in RLWE operations)
- $p$ (group modulus) = A 2048-bit prime

With $\log_2(2048) \approx 11$, we consider $\log p \approx 11$ in our calculations. Using these, we evaluate complexities as:

- Hashing: $\Theta(k) = 256$
- Participants iteration: $\Theta(n) = 10$
- Polynomial tasks: $\Theta(M \log M) = 1024 \times 10 = 10240$
- Multiple exponentiations: $\Theta(n \times \log p) = 10 \times 11 = 110$
- Singular exponentiation: $\Theta(\log p) = 11$

Applying a logarithmic scaling, we deduce:

- Hashing: $\log_2(256) \approx 8$
- Participants iteration: $\log_2(10) \approx 3.3$
- Polynomial tasks: $\log_2(10240) \approx 13.3$
- Multiple exponentiations: $\log_2(110) \approx 6.8$

- Singular exponentiation: $\log_2(11) \approx 3.5$

The approximations based on our analysis are delineated in Fig. 2 with a comprehensive summary table.

Table 1. Summary of NIST SP800-22 Test Suite Results

| Test Case Name | Total Tests | Average P-Values | Pass | Fail | Pass % |
|---|---|---|---|---|---|
| Approximate Entropy Test | 16 | 0.3929 | 16 | 0 | 100.0 |
| Frequency Test within a Block | 16 | 0.4246 | 16 | 0 | 100.0 |
| Cumulative Sums Test | 16 | 0.4107 | 16 | 0 | 100.0 |
| Discrete Fourier Transform (Spectral) Test | 16 | 0.5096 | 16 | 0 | 100.0 |
| Frequency (Monobit) Test | 16 | 0.4883 | 16 | 0 | 100.0 |
| Linear Complexity Test | 16 | 0.6370 | 15 | 1 | 93.75 |
| Test for the Longest Run of Ones in a Block | 16 | 0.5632 | 16 | 0 | 100.0 |
| Non-overlapping Template Matching Test | 16 | 0.8817 | 16 | 0 | 100.0 |
| Overlapping Template Matching Test | 16 | 0.6907 | 13 | 3 | 81.25 |
| Runs Test | 16 | 0.4968 | 16 | 0 | 100.0 |
| Serial Test | 16 | 0.5087 | 16 | 0 | 100.0 |
| **Total** | 176 | 0.5459 | 172 | 4 | 98.86 |

## 6 Evaluation and Deployment

### 6.1 Entropy Estimation

For our MPC-based VRF scheme, we approximate the generated randomness as a random variable $X$ ranging $[0, 2^{256} - 1]$. The entropy of $X$ is computed considering its probability distribution function (PDF):

1. The seed, considered uniformly distributed over $[0, 2^{256} - 1]$, produces a Ring-LWE encryption [21]. This encryption's PDF mirrors the seed's.
2. The Ring-LWE encryption, derived deterministically from the seed, retains this PDF.
3. Randomness is crafted using a distributed MPC [44]. It's shaped as a weighted summation of participant commitments, with weights guided by participant shares.
4. Each commitment, deemed uniformly distributed over $[0, 2^{256} - 1]$, is independent, providing a consistent PDF.
5. Participant shares, static and independent, yield a delta-function PDF at their value.
6. Summation weights are normalized, ensuring randomness stays within $[0, 2^{256} - 1]$. The resulting PDF is a truncated sum of commitments, cut-off at $2^{256} - 1$.

Denote commitment and share of participant $i$ as $C_i$ and $S_i$ respectively. $R$ symbolizes the randomness from the VRF system, with $n$ as the maximal participant count in the VRF MPC contract. Then, the PDF of $R$ is given by:

$$PDF_R(r) = \frac{1}{Z} \int_0^{2^{256}-1} \left( \prod_{i=1}^{n} PDF_C(C_i) \right) \times \delta\left( r - \frac{\sum_{i=1}^{n} C_i \cdot S_i}{\sum_{j=1}^{n} S_j} \right) U(0, 2^{256} - 1)(r) dr \quad (3)$$

where $U(a, b)$ is the uniform distribution over the range $[a, b]$, and $Z$ is the normalization constant given by:

$$Z = \int_0^{2^{256}-1} \left( \prod_{i=1}^{n} PDF_C(C_i) \right) U(0, 2^{256} - 1) \times \left( \frac{\sum_{i=1}^{n} C_i \cdot S_i}{\sum_{j=1}^{n} S_j} \right) dr \quad (4)$$

To calculate the entropy of our hybrid VRF system, we first need to calculate the Shannon

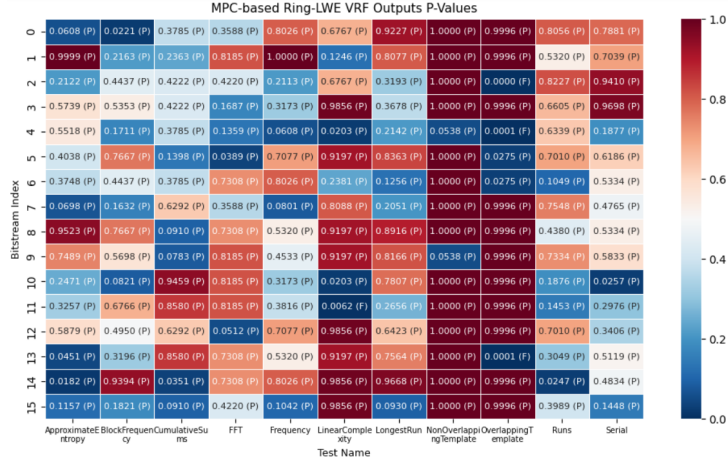

**Fig. 3.** *P*-value results from the NIST SP800-22 Test Suite

entropy [45] of the $PDF_R(r)$ formula. The Shannon entropy is given by the following formula:

$$H = -\int_0^{2^{256}-1} PDF_R(r) \log_2(PDF_R(r)) dr \quad (5)$$

Using the definition of $PDF_R(r)$, we have:

$$H = -\int_0^{2^{256}-1} \left[ \frac{1}{Z} \int_0^{2^{256}-1} \left( \prod_{i=1}^n PDF_C(C_i) \right) \times \delta\left( r - \frac{\sum_{i=1}^n C_i \cdot S_i}{\sum_{j=1}^n S_j} \right) U(0, 2^{256}-1)(r) dr \right] \\ \times \log_2 \left[ \frac{1}{Z} \int_0^{2^{256}-1} \left( \prod_{i=1}^n PDF_C(C_i) \right) \times \delta\left( r - \frac{\sum_{i=1}^n C_i \cdot S_i}{\sum_{j=1}^n S_j} \right) U(0, 2^{256}-1)(r) dr \right] dr \quad (6)$$

This is a complex expression due to the nested integrals and delta function. Now, making use of the properties of the delta function:

$$\int_0^{2^{256}-1} f(r) \delta(r-a) dr = f(a) \quad (7)$$

We can simplify our expression for entropy. Additionally, as the commitments $C_i$ are uniformly distributed over $[0, 2^{256}-1]$, their PDF is:

$$PDF_C(C_i) = \frac{1}{2^{256}} \quad (8)$$

Substituting this in, the entropy formula simplifies further:

$$H = -\int_0^{2^{256}-1} \left( \frac{1}{Z} \left( \frac{1}{2^{256}} \right)^n U(0, 2^{256}-1)(r) \right) \times \log_2 \left( \frac{1}{Z} \left( \frac{1}{2^{256}} \right)^n U(0, 2^{256}-1)(r) \right) dr \quad (9)$$

Given that $U(0, 2^{256}-1)(r) = 1$ for $r$ in $[0, 2^{256}-1]$, this can be further simplified to:

$$H = -\left( \frac{1}{Z} \left( \frac{1}{2^{256}} \right)^n \right) \log_2 \left( \frac{1}{Z} \left( \frac{1}{2^{256}} \right)^n \right) \times 2^{256} \quad (10)$$

This is the final specific formula for the estimated entropy of $H$.

## 6.2 Randomness Evaluation

We used the NIST SP800-22 [22] test suite to evaluate the randomness attributes of binary sequences produced by our MPC-based Ring-LWE VRF system. From the suite, we utilized 11 out of the 15 tests, excluding the 'Binary Matrix Rank Test', 'Maurer's Universal Statistical Test', 'Random Excursions Test', and 'Random Excursions Variant Test' due to our data pattern discrepancies.

As summarized in Table 1 and Fig. 3, the system's performance is exemplary with the majority of tests achieving average $p$-values centered around 0.5, which signifies optimal randomness. Notably, the 'NonOverlappingTemplate' test yielded an exceptional average $p$-value of 0.881706.

Out of 176 tests, only 4 failed, marking an impressive pass rate of about 98.86%. The 'LinearComplexity' and 'OverlappingTemplate' tests recorded one and three failures, respectively. However, these failures merely suggest certain statistical patterns in isolated cases rather than a system-wide randomness flaw.

To sum up, our MPC-based Ring-LWE VRF system showcases strong randomness as validated by the NIST SP800-22 results, underscoring its reliability for applications necessitating quality random sequences.

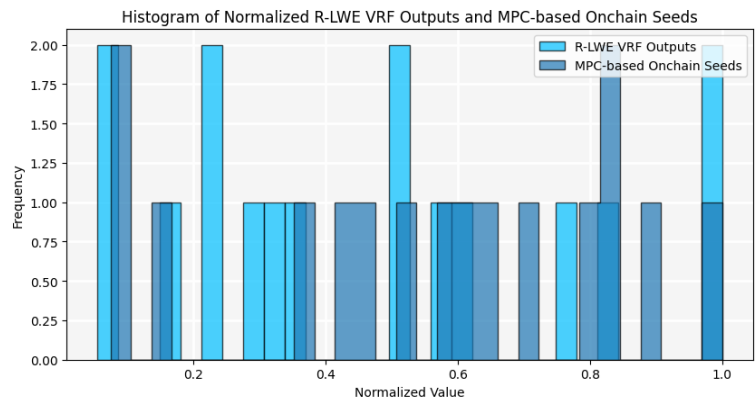

**Fig. 4.** Uniformly Distributed VRF outputs. The normalized values were derived from the raw 256-bit values.

## 6.3 System Deployment

We used Ganache to deploy our hybrid VRF system based on Truffle and Remix environment, which is a personal blockchain that allows to test and deploy smart contracts on a local network without incurring the cost and time delay associated with deploying on the main Ethereum network. The unique contract address, comprised of 160 bits, is represented by the hexadecimal value "0xd914 5CCE 52D3 86f2 5491 7e48 1eB4 4e99 43F3 9138". Concurrently, the contract creation transaction hash is given by "0x6e8c 5bda 09f1 4004 b75a 02b4 14fd 05c8 def8 70d8 0e65 d029 7dcc 364c 059b b9db". The execution of the creation transaction necessitated the expenditure of 3,425,664 gas units. Fig. 4 provides a visual representation of the output VRF distribution, where the uniform distribution property is observed for both the MPC-based VRF on-chain seeds and the final R-LWE VRF outputs. Fig. 5 shows more characterized uniform distribution of ones in histogram along with scattered dispersion of the generated MPC-based seeds and Ring-LWE VRF outputs. Fig. To illustrate as an example, the 256-bit VRF output value was "0x46a6 f730 ead6 a473 1204 e2d9 e96c e2ed 9b5c d9df 0cda e614 bde6 e986 c68e 9ccd" when the MPC-based on-chain seed was given as "0x9a23 982c 68ed 7fa1 4863 a3b1 d796 22fe 6f7b 9a28 fd06 dd1c 41d0 7f1b b9c1 708e".

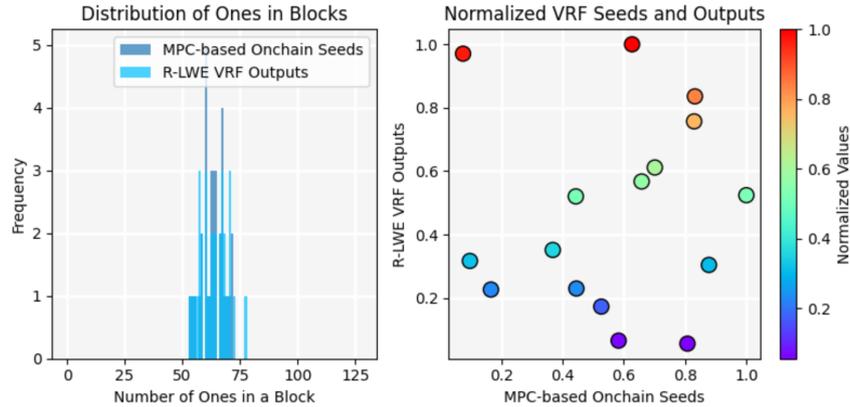

**Fig. 5.** The distribution of ones in 128-bit blocks and scattered distribution of VRF seeds and output. The average ratios of ones were observed as 0.4912 and 0.4934 in the MPC-based seeds and Ring-LWE VRF output, respectively.

## 7 Conclusion

This paper presents a blockchain-based VRF scheme to address quantum threats to traditional cryptographic protocols, leveraging post-quantum Ring-LWE encryption. To tackle the computational overhead and gas costs of Ring-LWE, our architecture seamlessly merges on-chain and off-chain computations. Off-chain operations are verified via a DID-based ring signature with a Non-Interactive Zero-Knowledge (NIZK) proof and delegated key generation, inspired by the Chaum-Pedersen proof and Fiat-Shamir Heuristic. By integrating multi-party computation (MPC) and blockchain-based decentralized identifiers (DID), our VRF enhances randomness while bolstering security and privacy. Evaluations highlight the model's robustness, evidenced by its 98.86% pass rate on 11 standard tests from the NIST SP800-22 suite and an average $p$-value of 0.5459 across 176 tests. The results validate the VRF's theoretical rigor and practical suitability for scenarios demanding verifiable randomness.

## Acknowledgment

This research is supported by SUNY Korea and the Macao Polytechnic University research grant (Project code: RP/FCA-02/2022) and the National Research Foundation of Korea (NRF) grant funded by the Ministry of Science and ICT (MSIT), Korea (No. 2020R1F1A1A01070666).

## Authors

**Bong Gon Kim** earned his B.S. in Electrical & Electronics and M.S. in Computer Science from Yonsei University. He's a Ph.D. candidate at the Stony Brook University's Computer Science Department. He served as a Senior Engineer at Samsung Electronics' SW Platform Lab from 2003 to 2018, contributing to mobile communication projects, Android apps, and Linux systems. His research now centers on DID-based identity, cryptographic security, ring signatures with NIZK proofs, and privacy in blockchain systems.

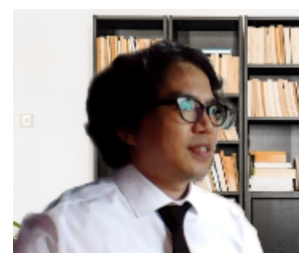

**Dennis Wong** secured his Ph.D. in Computer Science from the University of Guelph and an undergraduate degree from the Chinese University of Hong Kong. Guided by combinatorics, string algorithms, and graph theory, his studies span Gray codes, de Bruijn sequences, and algorithm design and analysis. Lately, he has developed interests in financial modeling and sports analytics.

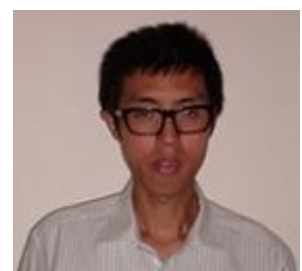

**Yoon Seok Yang** completed his B.S. and M.S. degrees at Hanyang University and later acquired another M.S. degree from the University of California, Irvine. He achieved his Ph.D. in Electrical and Computer Engineering from Texas A&M University. Between 2000 and 2005, he contributed as a Research Engineer at LG Electronics. Now with Intel Corporation, his focus lies in neuromorphic computing, SoC designs for AI, NoC designs for multiprocessor setups, and edge computing.

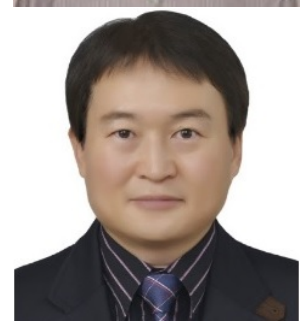